\documentclass[twocolumn,showpacs,prl]{revtex4}

\usepackage{graphicx}%
\usepackage{dcolumn}
\usepackage{amsmath}
\usepackage{latexsym}

\begin{document}

\title{Growing smooth interfaces with inhomogeneous, moving 
external fields: dynamical transitions, devil's staircases and  
self-assembled ripples}

\author{Abhishek Chaudhuri, P. A. Sreeram and Surajit Sengupta\\
}
\affiliation{
Satyendra Nath Bose National Centre for Basic Sciences, \\
Block-JD, Sector-III, Salt Lake,
Calcutta - 700098.
}

\date{\today}

\begin{abstract}
\noindent
We study the steady state structure and dynamics of an interface
in a  pure Ising system on a square  lattice 
placed in an {\em inhomogeneous} external field with a profile designed 
to stabilize a flat interface, and  
translated with velocity $v_{e}$. For small $v_{e}$, the interface is stuck 
to the profile, is macroscopically smooth, 
and is rippled with a periodicity in general incommensurate with the 
lattice parameter. For arbitrary orientations of the profile, the local slope
of the interface locks in to one of infinitely many rational values 
(devil's staircase) which most closely approximates the profile. 
These ``lock-in'' structures and ripples dissappear as $v_e$ increases. 
For still larger
$v_{e}$ the profile detaches from the 
interface.  
\end{abstract}
\maketitle

The ability to grow flat solid surfaces\cite{magnet} is often 
of major technological concern, for example, in the fabrication of 
magnetic materials 
for recording devices where surface roughness\cite{degrade} causes a sharp
deterioration of magnetic properties. Most growing surfaces or interfaces
on the other hand coarsen\cite{Barabasi} with a width, $\sigma$, 
which diverges with system size as $L^{\alpha}$ and time as  $t^{\beta}$,
where $\alpha$ and $\beta$ ($\alpha, \beta \ge 0$) are the 
{\em roughness}  and  {\em dynamical} exponents respectively. 
Is it possible to drive such an interface with a 
pre-determined velocity $v_f$ and, at the same time, keep
it flat (i.e. $\alpha = \beta = 0$)\,?
In this Letter, we explore this possiblity by studying growing interfaces 
in a non-uniform field with an appropriately shaped profile 
moving without change of shape at an 
externally controllable velocity $v_e$. We find that for $v_e$ less than a 
limiting value $v_{\infty}$, it is possible to produce a macroscopically flat,
interface growing with average velocity $v_f = v_e$. Microscopically, however,
the interface shows an infinity of {\em dynamical} ripple structures
similar to self-assembled, commensurate~-incommensurate(C-I)
\cite{FK} domains produced by atomic mismatch\cite{mismatch}.
The ripples vanish with increasing $v_e$ through a
{\em fluctuation induced} C-I transition.

Real solid-solid interfaces being complex\cite{sutton}, an understanding
of the dynamics of such interfaces in a general 
time~-dependent, inhomogeneous field may only be achieved by 
beginning with a relatively simple, but nontrivial, system viz. an
interface in the two~-dimensional (2-d) Ising model. Our results, therefore,
concern mainly this model system, though towards the end we discuss
possible modifications, if any, for solid interfaces.

The dynamics of an Ising interface in a (square) lattice driven by 
uniform external fields is a rather well studied\cite{Barabasi} 
subject. The velocity, $v_{\infty}$ of the interface depends on the applied 
field, $h$ and the orientation $\theta$ measured with respect to the underlying 
lattice. The interface is rough and coarsens with KPZ\cite{Barabasi} exponents 
$\alpha = 1/2$ and $\beta= 1/3$. 

\begin{figure}[t]
\begin{center}
\includegraphics[width=6cm,height=9.2cm]{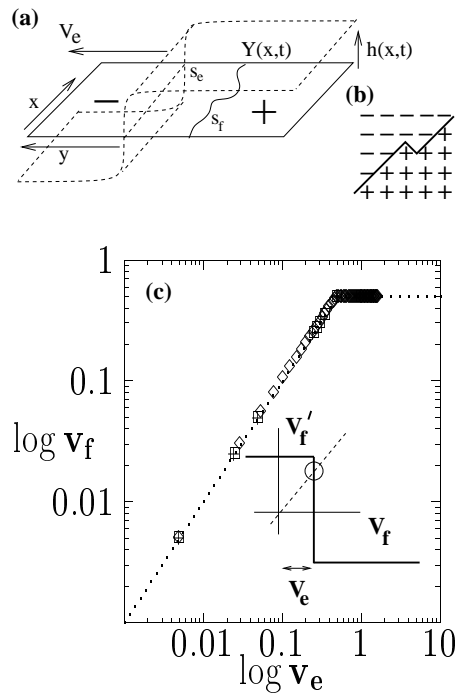}
\end{center}
\caption{
(a) An Ising interface $Y(x,t)$ (bold curved line) between regions of positive 
(marked $+$) and
negative (marked $-$) magnetization in an external, inhomogeneous field 
with a profile which is as shown(dashed line). The positions of the edge of the
field profile and that of the interface are labelled $S_e$ and $S_f$ respectively.
(b) A portion of the interface in a square lattice showing a corner. 
(c) The interface velocity $v_f$ as a function of the velocity of the
dragging edge $v_e$ for $N_s = 100(\Box), 1000(\Diamond), 10000(+)$
and $\rho = 0.5$. All the data ($\Box,\Diamond,+$) collapse on the
mean field solution (dashed line). Inset shows the graphical solution
(circled) of the self-consistency equation for $v_f$; dashed line
represents $v_f = v_f$.
}
\label{modvel}
\end{figure}

Consider, an interface $Y(x,t)$ between phases with 
magnetization, $\phi(x,y,t) > 0$ and $\phi(x,y,t) < 0$, in a 2-d square 
lattice\cite{mexpln} obeying single-spin flip Glauber 
dynamics \cite{Men} in the limit $h/J , T/J \rightarrow 0$. 
Here $J$ is the Ising exchange coupling and $T$ the temperature.
An external non-uniform field  with a profile 
$h(y,t) =  h_{max} f(y,t)$ where $f(y,t) = \tanh ((y - v_{e}t)/\chi)$ 
and $\chi$ is the
width of the profile (see Fig~\ref{modvel}(a)) is applied
such that $h = {\it h_{max}}$ in the  $\phi > 0$ 
and $-{\it h_{max}}$ in the $\phi < 0$ regions separated by
a sharp {\it edge}. The driving force 
depends on the relative {\em local} position of 
$Y(x,t)$ and the edge. In the low temperature limit the 
interface moves solely by random corner flips\cite{Barabasi} 
(Fig.~\ref{modvel}(b)), the fluctuations  
necessary for nucleating islands of the minority phase in any region being
absent. We study the behaviour of $v_f$ and the 
structure of the interface as a function of $v_e$ and orientation.

Naively, one expects fluctuations of the interfacial coordinate $Y(x,t)$
to be completely suppressed in the presence of $h(y,t)$.
Indeed, as we show below, a mean field theory obtains 
the exact behaviour of $v_f$ as a function of $v_e$
(Fig. \ref{modvel} (c)). 
Using model A dynamics\cite{Chaikin} for a coarse-grained Hamiltonian 
(which, for the moment, ignores the lattice) of an 
Ising system in a external non-uniform field   
together with the assumption that 
the magnetization $\phi$ is uniform everywhere except near the interface, 
one can derive\cite{physica} an
equation of motion for the interface. 
\begin{equation}
\frac{\partial Y}{\partial t} = \lambda_{1} \frac{\partial^{2} Y}
{\partial x^{2}}-\lambda_{2} \Big(\frac{\partial Y}{\partial x} \Big)^{2} 
f(Y,t)-\lambda_{3} f(Y,t)+ \zeta(x,t)
\label{m-KPZ}
\end{equation}
where $\lambda_{1}$,$\lambda_{2}$ and $\lambda_{3}$ are parameters 
and $\zeta$ is a Gaussian white noise with zero mean.
Note that Eq.\,(\ref{m-KPZ}) lacks Galilean invariance\cite{multif} 
$Y^{\prime} \rightarrow Y + \epsilon x,\>\> x^{\prime} \rightarrow x - 
\lambda_{2}\epsilon t, \>\> t^{\prime} \rightarrow t$. 
A mean field calculation amounts to taking $Y \equiv Y(t)$ i.e. neglecting 
spatial fluctuations of the interface and noise. 
For large times ($t \rightarrow \infty$), $Y \rightarrow v_{f}t$, where 
$v_{f}$ is obtained by solving the self-consistency equation; $
v_{f} = \lim_{t\to \infty} -\lambda_{3}\tanh\{(v_{f} - v_{e})t/\chi\}
      = -\lambda_{3}\,{\rm sign}(v_{f} - v_{e})
$.
For small $v_e$ the only solution (Fig.~\ref{modvel}(c), inset) is 
$v_f = v_e$ and for  $v_{e} > v_{\infty}$, where 
$v_{\infty} = \lambda_{3}$ we get 
$v_{f} = \lambda_{3} = v_{\infty}$. We thus have a sharp transition\, 
(Fig.~\ref{modvel}(c)) from a region where the interface is stuck to the edge 
to one where it moves with a constant velocity.
How is this result altered by including spatial fluctuations of $Y$ ?
This question is best answered by mapping the interface problem to a 
1-d cellular automaton\cite{Barabasi,AEP}.

The interface coordinate $Y(x,t)$ in a square lattice is 
represented\cite{AEP,Maj1} by the set of integers
$\{y_i\},\,\,1<\,i\,<\,N_p$ denoting positions of $N_p$ hard-core 
($y_{i+1} \geq y_i + 1$) particles in a 1-d lattice of $N_s$ sites.
The particle density $\rho = N_p/N_s$ determines the mean slope
of the interface $\tan \theta_f = 1/\rho$ and the motion of the 
interface by corner flips corresponds to   
the hopping of particles with right and left jump probabilities 
$p$ and $q$ ($p+q = 1$). Trial moves are attempted sequentially on randomly
chosen particles\cite{AEP} and $N_p$ attempted hops constitute a single 
time step. In our case, $p$ and $q$ are position dependent such that,
$\Delta_i(v_e\,t) = p-q = \Delta\,{\rm sign}(y_i- i/\rho - v_e t)$
with $\Delta = 1$. The bias $\Delta_i(v_e\,t)$ is 
appropriate for a step function ($\chi = 0$) profile with  
the slope of the profile edge equal to the average slope of 
the interface. We track the instantaneous particle velocity $v_f(t)$ defined
as the number of particles moving right per unit time,
the average position $<y(t)> = N_p^{-1}\sum_{i=1,N_p}y_i(t)$, the 
width 
$\sigma^{2}(t)=N_p^{-1}\sum_{i=1,N_p}<(y_i(t)\,-<y_i(t)>)^2>$
and the local slope of the interface given by the local density of particles. 
Angular brackets denotes an average over the realizations of the 
random noise. The usual particle hole symmetry for an exclusion 
process\cite{Barabasi,AEP,Maj1}
is violated since exchanging particles and holes alters the relative position
of the interface compared to the edge.

We perform numerical simulations\cite{AEP} of the above 
model for $N_s$ upto $10^4$ to obtain $v_f$ for the steady 
state interface as a function of $v_e$ as shown in Fig.~\ref{modvel}(c).  
A sharp dynamical transition from an initially stuck interface with 
$v_f = v_e$ to a free, detached interface with 
$v_f = v_\infty = \Delta (1-\rho)$ is clearly evident as predicted 
by mean field theory. The detached interface coarsens with KPZ 
exponents\cite{physica}.
Note that, even though the mean field solution for $v_f(v_e)$ neglects 
the fluctuations present in our simulation, it is exact. The detailed 
nature of the stuck phase 
($v_f = v_e$ and $\sigma$ bounded) is, on the other hand,
considerably more complicated than the mean field assumption $Y(x,t)=Y(t)$.

\begin{figure}[t]
\begin{center}
\includegraphics[width=8cm]{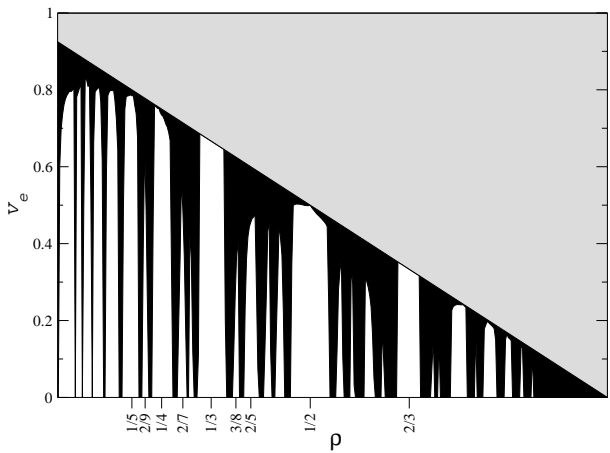}
\end{center}
\caption[]
{The dynamical phase diagram in $v_e$ and $\rho$ plane. The numbers on the 
$\rho$ axis mark the fractions $\tilde{\rho}$, which determines
the orientation of the lock-in phase. The three regions white, black and grey 
correspond to the rippled, the disordered and the detached 
phases respectively.} 
\label{pdia}
\end{figure}

\begin{figure}[t]
\begin{center}
\includegraphics[width=8cm]{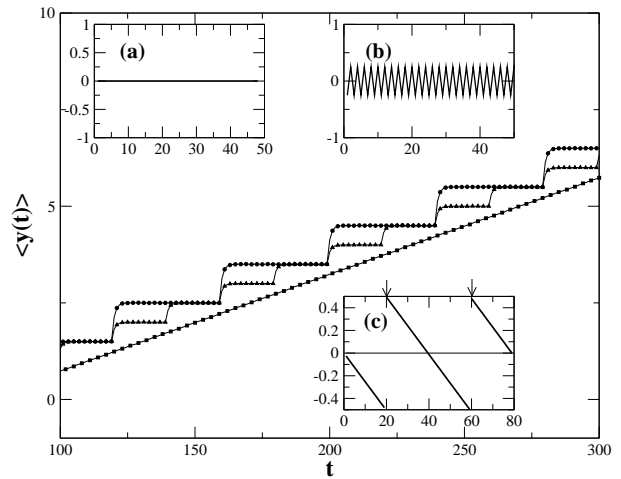}
\end{center}
\caption{ Variation of $<y(t)>$ with $t$ for 
$v_{e} = 0.025$ and $p = 1.0$. Lines denote analytic
results while points denote Monte Carlo data for $\rho = 1/5$ (uppermost curve),
$2/5$ and an incommensurate $\rho$ near $1/3$.
Inset (a)-(c) shows the corresponding ground state 
interfaces ($y_i-i/\rho$). The arrows in (c) mark the positions of two 
discommensurations.
}
\label{hops}
\end{figure}

The ground state of the interface in the presence of a stationary 
($v_e = 0$) field profile is obtained by minimizing 
$E = 1/N_p\sum_i (y_i- i/\rho - c)^2$ with respect to the set $\{y_i\}$ 
and the constant, $c$, subject to the constraint that $y_i$ are integers. 
The form of $E$ leads 
to an additional  non-local, repulsive, interaction between particles. 
The minimized energy may be calculated  
exactly, $E(\rho = m/n) = \frac{1}{6}(\frac{1}{2}-\frac{1}{m})(1-\frac{1}{m})+
\frac{1}{4m}-\frac{1}{4m^2}$ for $m$ {\em even} and $\frac{1}{12}(1-\frac{1}{m^2})$ for $m$ 
{\em odd}, where the density $\rho = m/n$, is a rational fraction. 
The energy satisfies the bounds $E(1/n) = 0 < E(m/n) < \lim_{m\to \infty}E(m/n) = 1/12$ where the upper bound is for an irrational density.
For an arbitrary $0\,<\,\rho\,<\,1$ the system ($\{y_i\}$) prefers 
to distort, conforming within local regions, to the nearest low-lying 
rational slope $1/\tilde{\rho}$ interspersed with an {\em ordered} 
array of ``discommensurations''
of density $\rho_d = |\rho - \tilde{\rho}|$ which appear as long wavelength 
{\em ripples} (see Fig.~\ref{hops} inset (c)).  
The $\tilde{\rho}$ as a function of $\rho$ shows a ``devil's staircase'' 
structure (complete for $v_e \to 0$) with a multifractal\,\cite{CDW} measure.
We observe this in our simulation by 
analyzing the instantaneous distribution of the local density of 
particles to obtain weights for various simple rational fractions
upto generation $g=9$ in the Farey tree of rationals\cite{farey}.  
A time average of the density corresponding to the fraction with 
largest weight at any $t$,
then give us the most probable density
$\tilde{\rho}$ --- distinct from the average $\rho$ which is constrained 
to be the inverse slope of the profile. For small $v_e$ the interface is 
more or less locked in to a single $\tilde{\rho}$, shown as white regions 
in the phase diagram (Fig.~\ref{pdia}) in the $v_e - \rho$ plane.  

For low velocities and density where correlation effects due to the 
hard core constraint are negligible, the dynamics of the interface may be 
obtained exactly\cite{physica}. Under these circumstances the $N_p$ 
particle probability distribution for the $y_i$'s, $P(y_1,y_2,\cdots,y_{N_p})$
factorizes into single particle terms $P(y_i)$. Knowing the time development
of $P(y_i)$ and the ground state structure the motion of 
the interface at subsequent times may be trivially computed as a sum of 
single particle motions. 
A single particle (with say index $i$) moves with
the bias $\Delta_i(v_e\,t)$ which, in general, may change sign at 
$y < i/\rho+v_e\,t < y + 1$. Solving the appropriate set of master 
equations\cite{physica} we obtain,
for $v_e << 1$ the rather obvious steady state solution 
$P(y_i) = 1/2(\delta_{y_i,y}+\delta_{y_i,y+1})$ and the particle 
oscillates between $y$ and $y+1$. Subsequently, when $i/\rho+v_e\,t \geq y+1$,
the particle jumps to the next position and $P(y_i)$ relaxes exponentially
with a time constant $\tau = 1$ to it's new value with $y \to y+1$.
In general, for rational $\rho = m/n$, the motion
of the interface is composed of the independent motions of $m$ particles
each separated by a time lag of $\tau_{L} = 1/m\,v_e$.
The result of the analytic calculation for small $v_e$ and $\rho$ has been 
compared to those from simulations in Fig.~\ref{hops}  for 
$\rho = 1/5$  and $2/5 $. 
For a general irrational $\rho < 1/2$, $m \to \infty$ consequently, 
$\tau_{L} \to 0$. The 
$y_i$'s are distributed uniformly around the mean implying $\sigma^2 = 1/3$
independent of system size and time.
For $\rho > 1/2$ the width $\sigma^2 = (1-\rho)/3\rho$
since the number of mobile particles 
decreases by a factor of $(1-\rho)/\rho$. 

The forward motion of an irrational interface is accompanied by the motion of 
discommensurations along the interface with velocity $v_e$ which 
constitutes a kinematic wave\cite{Barabasi,Maj1} parallel to the interface. 
As the velocity $v_e$ is increased 
the system finds it increasingly difficult to maintain
its ground state structure and for $\tau \geq \tau_L$
the instantaneous value of
$\tilde{\rho}$ begins to make excursions to other nearby low-lying fractions
and eventually becomes free. 
Steps corresponding to ${\tilde \rho}= m/n$ dissappear (i.e. 
${\tilde \rho}\to\rho$)
sequentially in order of decreasing $m$ and the interface loses the ripples.
The interface is disordered though $\alpha$ and $\beta$ continue to be zero
(black region in Fig.~\ref{pdia}). 
The locus of the discontinuities (within an accuracy of $1/N_p$)
in the $\tilde{\rho}(\rho)$ curve for various velocities $v_e$ gives the
limit of stability of the lock-in rippled phases.

\begin{figure}[t]
\begin{center}
\includegraphics[width=6.8cm]{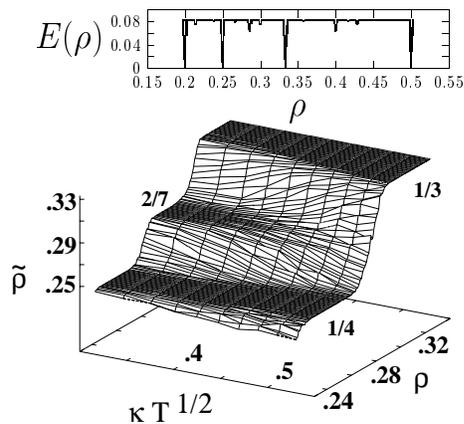}
\end{center}
\caption{
A surface plot of $\tilde{\rho}$ vs $\rho$ and $\sqrt{T} \kappa$ showing 
steps for the fractions $1/4,2/7$ and $1/3$. Note that the 
step corresponding to $2/7$ 
vanishes at $\sqrt{T} \kappa > 0.5$. Inset (top) shows a 
plot of $E(\rho)$ for a $N_s = 420$ site system.
}
\label{lang}
\end{figure}

While the stability of mismatch domains\cite{mismatch} is 
decided, mainly, by competition between mechanical, long-ranged 
(elastic) and short-ranged (atomistic) interactions\cite{FK}, dynamical 
ripples vanish with increasing $v_e$ through increased fluctuations. 
We argue that it is sufficient to project the 
entire configuration space ${y_i}$ of the stuck interface onto the single 
variable $\rho$. In Fig.~\ref{lang} (inset) we have plotted the energy 
$E(\rho)$ for the ground-state configuration with density $\rho$. 
As is obvious from the 
figure, $E(\rho)$ has a structure similar to the free energy surface of a 
1-d ``trap'' model\cite{trap} often used to describe glassy dynamics. 
The distinction, of course, is the fact that the energies of the traps in 
this case are highly correlated. We may 
then describe the dynamics of the stuck interface as the Langevin 
dynamics\cite{Chaikin} of a single particle with coordinate 
$\rho^\prime$ diffusing on a energy surface given by,
$
F(\rho^\prime) = E(\rho^\prime) + \kappa\,(\rho^\prime - \rho)^2,
$
kicked by a Gaussian white noise of strength $T$.
The second term, containing the modulus $\kappa$, 
ensures that $\rho^\prime \to \rho$ for $t \to \infty$. 
At intermediate times, however, 
the system may get trapped indefinitely in some nearby low-lying minimum 
with $\rho^\prime = \tilde{\rho}$ if $T$ ($\propto v_e^2$
by symmetry) is not large 
enough. As $T$ increases, the time spent in  jumping between minima may 
exceed the  residence time in the minimum resulting in 
a noise induced C-I transition (Fig.~\ref{lang}) from $\tilde{\rho}$.
 
In this Letter, we have studied the static and dynamic properties of an 
Ising interface in 2-d subject to a non-uniform, time-dependent external 
magnetic field. The system has a rich dynamical phase diagram with 
infinitely many steady states. The nature of these steady 
states and their detailed dynamics depend on the orientation of the 
interface and the velocity of the external field profile.
How are our results expected to change for real driven solid interfaces?
Firstly, real field profiles
would have a finite width $\chi > 0$. However, as long as $\chi$ is comparable to
atomic dimensions we find no appreciable change in the results. 
Secondly, elastic interaction between ``particles'' or 
steps in the interface\cite{sutton} may 
smoothen the devil's staircase though we expect that for 
typical solids this will be minimal. 
Thirdly, the structure of the underlying lattice may influence the stability of 
particular orientations.
Finally, exciting new physics may come into play as new modes e.g.
point and line defects\cite{mart}, as well as phonon degrees of 
freedom (leading to acoustic emmissions\cite{sutton}) are accessed as $v_e$ 
increases.\\  
{\bf Acknowledgement: } The authors thank M. Barma, J. K. 
Bhattacharya, J. Krug, S. N. Majumdar, A. Mookerjee and M. Rao for useful discussions; 
A. C. thanks C.S.I.R., Govt. of India for a fellowship.

 

\end{document}